\documentclass[prb,showpacs,twocolumn,superscriptaddress]{revtex4}
\usepackage{graphicx}
\begin{document}
\title{Superconducting Pairing Symmetries in Anisotropic Triangular Quantum Antiferromagnets}

%on an anisotropic triangular lattice for Layered Organic Conductors}

\author{J. Y. Gan$^1$, Yan Chen$^2$, and F. C. Zhang$^2$}

\affiliation{Center for Advanced Study, Tsinghua University, Beijing, 100084, China\\
$^2$Department of Physics and Center of Theoretical and
Computational Physics, The University of Hong Kong, Hong Kong,
China}

\begin{abstract}
Motivated by the recent discovery of a low temperature spin liquid
phase in layered organic compound $\kappa$-(ET)$_2$Cu$_2$(CN)$_3$
which becomes a superconductor under pressure, we examine the phase
transition of Mott insulating and superconducting (SC) states in a
Hubbard-Heisenberg model on an anisotropic triangular lattice.  We
use a renormalized mean field theory to study the Gutzwiller
projected BCS wavefucntions. The half filled electron system is a
Mott insulator at large on-site repulsion $U$, and is a
superconductor at a moderate $U$. The symmetry of the SC state
depends on the anisotropy, and is gapful with $d_{x^2-y^2}+id_{xy}$
symmetry near the isotropic limit and is gapless with $d_{x^2-y^2}$
symmetry at small anisotropy ratio.

\end{abstract}


%74.70.Kn Organic superconductors
%74.20.Rp Pairing symmetries (other than s-wave)
%74.25.Dw Superconductivity phase diagrams
\maketitle

Understanding of exotic properties in layered organic
superconductors $\kappa$-(ET)$_2$$X$ (X=anion) has attracted much
interest
recently~\cite{jerome,mckenzie97,ishiguro,lang,Kuroki,Nature05}. In
these materials, (ET)$_2$ dimers are arranged in a
quasi-two-dimensional (quasi-2D) anisotropic triangular lattice, as
schematically shown in Fig. 1(a). Since the intradimer hopping
integrals are much larger than the interdimer ones, and the carrier
density of the compound is one hole per dimer, the low energy
electronic structure can be well approximated by a 2D Hubbard model
at the half filling, where each lattice site represents a
dimer~\cite{fukuyama}, as illustrated in Fig. 1(b) with nearest
neighboring (n.n.) and next n.n. hopping integrals $t$ and
$t^\prime$ and a strong on-site Coulomb repulsion $U$. Under the
pressure, some of the compounds show a first-order transition from a
Mott insulator to a superconductor. We expect the effect of $t'$ or
the geometrical frustration to be crucial to the low temperature
phases. In most of $\kappa$-(ET)$_2$$X$ materials, $t'/t\sim
0.4-0.8$, the effect of geometrical frustration is less remarkable
so that the ground state of the insulating phase is
antiferromagnetic, similar to the high-$T_c$ SC
cuprates~\cite{miyagawa,mayaffre,soto,lefebvre,miyagawa1}. An
example is the compound with X=Cu[N(CN)$_2$]Br, where $t'/t \sim
0.6$. For systems with a larger $t'/t$, the frustration may destroy
magnetic ordering at low temperatures and the interplay between the
frustration and strong correlation may lead to the possible
emergence of spin liquid state or unconventional superconductivity.
The compound of $X=$Cu$_2$(CN)$_3$ ($t'/t \sim 1.06$) appears to be
an excellent candidate for spin liquid state without manifestation
of any magnetic ordering as indicated in the NMR at low temperatures
down to 32 mK~\cite{shimizu}, reminiscent of the resonating valence
bond state proposed by Anderson~\cite{anderson73,vannila} in a
triangular lattice system. The compound becomes SC under moderate
hydrostatic pressure with the maximum $T_c$ of 3.9 K.

The SC pairing symmetry in the layered organic conductors remains
controversial. Experimentally, the NMR
measurements~\cite{mayaffre,soto,Kanoda}, the angular dependent
scanning tunneling microscopy~\cite{arai} and the thermal
conductivity measurements in the vortex state~\cite{izawa} all
suggest the existence of nodes in the gap and indicate a spin
singlet pairing d$_{x^2-y^2}$ symmetry, similar to that of
high-T$_c$ cuprates. However, the temperature dependence of the
penetration depth\cite{Lang,Dressel} and of the specific heat
\cite{Elsinger} seem to support a full gapped SC state. Note that
there are also results consistent with a gap with
nodes\cite{Kanoda2,Carrington}. On the  theoretical side,
fluctuation exchange approximation for the Hubbard model shows a
pure $d_{x^2-y^2}$-wave symmetry~\cite{schmalian} while the
variational Monte Carlo and mean field study for a generalized
$t$-$J$ model ($t$-$t'$-$J$-$J'$ model) suggest that
$d_{x^2-y^2}+id_{xy}$- ($d+id$-) wave state is more
stable~\cite{ogata,ogata03,d+id}. Concerning the order of SC to Mott
insulator transition, there are two recent
studies~\cite{Liu05,Watanabe06} based on variational Monte Carlo
calculation by analyzing a 2D Hubbard model. They find a first-order
SC to Mott insulator transition for intermediate $U$ and the SC
state is restricted to the $d_{x^2-y^2}$-wave symmetry.

\begin{figure}
\includegraphics*[width=8.8cm,height=5.2cm,angle=0]{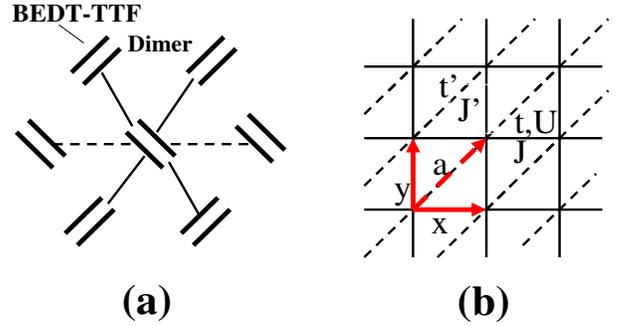}
\caption{(a) $\kappa-$type ET molecular arrangement within the
conducting plane. (b) Anisotropic triangular lattice structure of
the 2D Hubbard-Heisenberg ($t$-$t'$-$J$-$J'$-$U$) model. Also shown
are three unit vectors on the lattice.}
\end{figure}

In a previous paper, we considered the competition between
$d_{x^2-y^2}$-wave SC and antiferromagnetic states in
$\kappa$-(ET)$_2$ salts in the parameter region $t'/t <
0.8$~\cite{gan05}. The Gutzwiller renormalized mean-field
theory~\cite{zhang88} was employed to study an anisotropic
Hubbard-Heisenberg model in a 2D lattice. We found a first-order
transition between a Gossamer superconductor at a smaller $U$ and an
antiferromagnetic insulator at a larger $U$. Our result was
qualitatively consistent with major experiments. In this paper, we
are primarily interested in layered organic conductor with $t'/t
\sim 1$, with the compound $X=$Cu$_2$(CN)$_3$ in mind. Our focus
will be on the Mott insulator to SC transition and the pairing
symmetry of the SC state. We expect the anisotropic ratio $t'/t$
will affect the pairing symmetry. Since the compound of
$X=$Cu$_2$(CN)$_3$ is not magnetically long range ordered, we shall
neglect the antiferromagnetism in our study. We find that at half
filling, the ground state is a metal for small $U$, a Mott insulator
for large $U$, and a Gossamer superconductor at intermediate $U$,
whose pairing is $d+id$ near the isotropic point. A transition
between different pairing symmetries may show up by tuning the model
parameters.

We start with a Hubbard-Heisenberg model on an anisotropic
triangular lattice,
\begin{eqnarray}
H &=&U\sum_{i}n_{i\uparrow }n_{i\downarrow }-\sum_{\langle ij
\rangle\sigma }t_{ij}(c_{i\sigma }^{\dagger }c_{j\sigma }+
h.c.) \nonumber \\
&+& \sum_{\langle ij\rangle} J_{ij} S_{i}\cdot S_{j}
-\mu\sum_{i} n_{i},
\end{eqnarray}
where $c_{i\sigma }^{\dagger }$ is to create a hole with spin
$\sigma $ at site $i$, $S_{i}$ is a spin operator, $\mu$ is the
chemical potential, and $\langle ij \rangle$ denotes a neighboring
pair on the lattice. $t_{ij} = t$, $J_{ij}=J$ for the neighboring
pairs along $x$ and $y$ directions, and $t_{ij} = t'$, $J_{ij}=J'$
for the neighboring pairs along the $a$-direction as shown in Fig.
1(b). Hereafter we set $t=1$ and use $t$ as the energy unit, and
choose $J/t=1/3$ and consider mainly the system in the parameter
range $t^{\prime }/t=0.8-1.2$. We choose $J'/J=(t'/t)^{2} $,
consistent with the relation of $J= 4t^2/U$ in the large $U$ limit.
We will exclusively consider the half filled case. When
$t'/t=J'/J=1$, the model becomes isotropic. At $U \rightarrow
\infty$, the model is reduced to the Heisenberg model. We remark
that the Heisenberg model on an isotropic triangular lattice has a
long range antiferromagnetic order with an angle of $2\pi /3$
between neighboring spins. However, there is the possibility to have
a spin liquid state in the anisotropic Heisenberg
model~\cite{trumper} and in the anisotropic Hubbard
model~\cite{imada}. In what follows below, we will devote to the
non-magnetic states in our variational approach.

We consider a partially Gutzwiller projected BCS trial wavefunction
~\cite{laughlin,zhang03,Powell05},
\begin{eqnarray}
|\Psi _{GS}\rangle &=&\prod\limits_{i}P_i |\Psi _{BCS}\rangle   \nonumber\\
|\Psi _{BCS}\rangle &=&\prod\limits_{k}(u_{k}+v_{k}c_{k\uparrow
}^{\dagger }c_{-k\downarrow }^{\dagger })|0\rangle
\end{eqnarray}
where $P_i = z^{n_i}(1-\alpha n_{i\uparrow }n_{i\downarrow })$ is a
projection operator to partially project out the doubly occupied
electron states on the site $i$ and $0\leq \alpha \leq 1$ measures
the strength of the projection. $z$ is the fugacity to ensure the
charge density is unchanged by projection~\cite{laughlin,
anderson-ong,edegger,qhwang}. Note that introduction of the fugacity
does not change the previous results on the renormalized mean field
theory for the Gutzwiller state for that the number of charge
carriers were assumed to be the same~\cite{zhang88}.

We then apply the Gutzwiller approximation to estimate the
variational energy. In this scheme, the effect of the projection
operator is taken into account by a set of renormalized factors,
which are determined by statistical
countings~\cite{gutzwiller,vallhardt}. Let $\langle Q \rangle$ be
the expectation value of $Q$ in the state $|\Psi _{GS} \rangle$ and
$\langle Q \rangle_{0}$ be that in the state $|\Psi _{BCS} \rangle$.
The Gutzwiller approximation gives,
\begin{eqnarray}
\langle c_{i\sigma }^{\dagger }c_{j\sigma }\rangle \approx g_{t}\langle
c_{i\sigma }^{\dagger }c_{j\sigma }\rangle _{0} , ~~
\langle \vec{S}_{i}\cdot \vec{S}_{j}\rangle  \approx g_{s}\langle \vec{S}%
_{i}\cdot \vec{S}_{j}\rangle _{0}
\end{eqnarray}
where $g_{t}$ and $g_{s}$ are the Gutzwiller renormalized
coefficients and can be derived as follows~\cite{zhang03},
\begin{eqnarray}
g_{t} &=&\frac{(n-2d)(\sqrt{d}+\sqrt{1-n+d})^{2}}{(1-n/2)n}  \nonumber \\
g_{s} &=&[\frac{(n-2d)}{(1-n/2)n}]^{2}
\end{eqnarray}
where $n$ is the average electron number and $d=\langle
n_{i\uparrow}n_{i\downarrow} \rangle$ is the average electron
double occupation number. There is one-to-one correspondence
between $\alpha$ and $d$. Within the Gutzwiller approximation, we
have,
\begin{eqnarray}
(1 - \alpha)^2 = \frac{d(1-n+d)}{(n/2-d)^2}
\end{eqnarray}
In terms of the renormalized coefficients, we can get the effective
Hamiltonian,
\begin{eqnarray}
H_{eff}&=& Ud- g_{t}\sum_{\langle ij \rangle \sigma
} t_{ij} c_{i\sigma }^{\dagger }c_{j\sigma }+ h.c. \nonumber\\
&+& g_{s}\sum_{\langle ij \rangle} J_{ij} S_{i}\cdot S_{j}
-\mu \sum_{i}n_{i}.
\end{eqnarray}
The variation of the projected state $|\Psi _{GS} \rangle$
for $H$ in $(1)$ is thus reduced to the variation of the unprojected
state $|\Psi _{BCS} \rangle$ for $H_{eff}$.  Note that the Gutzwiller projection
on a BCS state may change the average number of electrons  of the state.

To proceed further, we introduce two types of mean fields, ($\tau=
\pm \hat{x}, \pm \hat{y},\pm \hat{a}$)
\begin{eqnarray}
\Delta_{i,i+\tau}&=&\langle c_{i\downarrow }c_{i+\tau \uparrow }-c_{i\uparrow
}c_{i+\tau \downarrow }\rangle _{0}\nonumber \\
\chi _{i,i+\tau}&=&\sum\limits_{\sigma}\langle c_{i\sigma }^{\dagger }c_{i+\tau\sigma }
\rangle _{0}.
\end{eqnarray}
We consider the translational invariant state with even parity,
$\Delta_{i,i+\tau} = \Delta_{i+\tau, i} =\Delta_{\tau}$, and  $\chi_{i,i+\tau}=\chi_{\tau}$.
Applying the mean field theory to $H_{eff}$, we obtain,
\begin{eqnarray}
|u_{\vec k}|^{2}&=& (E_{\vec k}+\varepsilon _{\vec k})/2E_{\vec k},\nonumber\\
\left\vert v_{\vec k}\right\vert ^{2}&=& (E_{\vec k}-\varepsilon _{\vec k})/2E_{\vec k}, \nonumber\\
E_{\vec k}&=&\sqrt{\varepsilon _{\vec k}^{2}+|F_{\vec k}|^{2}},
\end{eqnarray}
where
\begin{eqnarray}
\varepsilon_{\vec k} &=&  - g_{t} \sum_{\tau} 2 t_{\tau}\cos{k_{\tau}}
 -\mu -\frac{3g_{s}}{4} \sum_{\tau} J_{\tau} \chi _{\tau}\cos k_{\tau}  \nonumber\\
F_{\vec k} &=& \frac{3}{4}g_{s} \sum_{\tau} \Delta_{\tau}\cos {k_{\tau}}
\end{eqnarray}
where $k_{\tau} = \vec k \cdot \vec \tau$, and the sum in $\tau$
runs over the three unit vectors. In particular, we assume
$\Delta_a$ to be real, and $\Delta_x$ and $\Delta_y$ to have a phase
$\theta$ and $-\theta$ relative to $\Delta_a$, respectively
\begin{equation}
\Delta_{a}=|\Delta_a|,~ \Delta_{x}= |\Delta_x| e^{-i\theta},~
\Delta_{y}= |\Delta_y| e^{i\theta}.
\end{equation}
The case of $\Delta_{a}=0$ and $\theta=\pi/2$ corresponds to the
pure $d_{x^{2}-y^{2}}$ state. In the case $\theta =0$, we have an
extended $s$ wave state. In general, $\Delta _{a}\neq 0$ and $\theta
\neq \pi/2$, we have $d+id$ symmetry\cite{ogata03}.

The self-consistent equations are given as
\begin{eqnarray}
\Delta _{\tau }&=&\frac{1}{N}\sum_{\vec k} \cos {k_{\tau}}
 F_{\vec k}/ E_{\vec k} \nonumber
\\
\chi _{\tau}&=&-\frac{1}{N}\sum_{\vec k} \cos {k_{\tau}}
\varepsilon_{\vec k}/ E_{\vec k} \nonumber \\
\delta&=&\frac{1}{N}\sum_{\vec k}\varepsilon _{\vec k}/E_{\vec k}
\end{eqnarray}%
where the last equation is on the number of holes which is zero at
the half filled of our interest here, and $N$ is the number of total
lattice sites and the sum in $\vec k$ runs over the first Brillouin
zone. The ground state energy is
\begin{eqnarray}
E=Ud-2g_{t}\sum_{\tau}t_{\tau}\chi _{\tau}
 -\frac{3g_s}{8} \sum_{\tau} J_{\tau}(|\Delta _{\tau}|^{2}+ |\chi _{\tau}|^{2}).
\end{eqnarray}
All the mean fields are
determined self-consistently for each set of model parameters $U$, $t'/t$ and
a given variational parameter $d$, from which
we obtain the lowest energy state with
respect to the variation of $d$.

\begin{figure}
\centerline{\includegraphics[width=6.8cm]{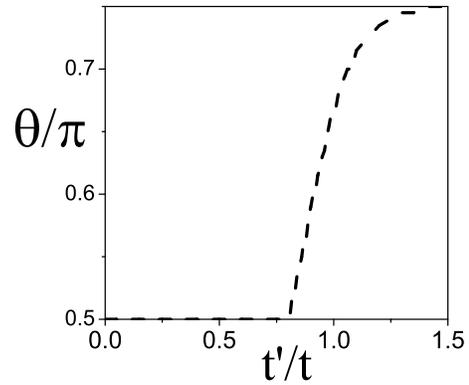}} \caption{The
relative phase of pairing parameter as a function of $t'/t$.}
\end{figure}

We first examine the evolution of the ground state by varying the
anisotropy ratio $t'/t$. The anisotropy ratio is a measure of the
geometrical frustration, which modifies the shape of Fermi surface
and therefore plays an important role in determining the competition
between Mott insulator and various SC states. As shown in our
previous study~\cite{gan05}, for small anisotropy $t'/t$, the ground
state is an antiferromagnetic insulator at large $U$ and a Gossamer
superconductor at medium $U$, followed by a normal metallic state at
small $U$, and the transition between Mott insulator and SC state
belongs to the first order. We expect the physics of the model in a
triangular lattice with large anisotropy may be qualitatively
different from that of a less anisotropic lattice. For large $t'/t$,
the antiferromagnetic state may be degraded into spin liquid state
and time-reversal symmetry broken SC state may show up as well. We
note that the transition from the Mott insulator to SC state remains
first order.

In our calculation, the relative phase $\theta$  of pairing
parameter is obtained by the minimization of the total energy. Due
to the numerical difficulty to determine the exact value of $\theta$
for small but finite $d$, we present the results in Fig. 2 for the
case $d=0$. The approximate critical value of $(t'/t)_c \sim 0.81$
distinguishes the different pairing symmetry region~\cite{footnote},
the SC order is always with $d_{x^{2}-y^{2}}$ symmetry
($\theta$=$\pi /2$) for $t'/t < (t'/t)_c$ while imaginary part of
pairing parameter emerges and the relative phase $\theta$ exceeds
$\pi/2$ for $t'/t > (t'/t)_c$. The SC order parameter with nonzero
imaginary part corresponds to $d+id$-wave state which breaks
time-reversal symmetry. The presence of such state may result in the
appearance of excitation gaps all over the Fermi surface. This
time-reversal symmetry broken state might be detectable by using
muon-spin rotation measurements~\cite{Pratt06}. It is interesting to
note that when $t'=t$ and $J'=J$, the relative phase $\theta$ is
equal to $2\pi /3$ where all the three amplitudes become symmetrical
and we have $\Delta$=$\Delta _{x}$=$\Delta _{y}$=$\Delta _{a}$ and
$\chi$ =$\chi _{x}$=$\chi _{y}$=$\chi _{a}$. This special pairing
symmetry is due to the geometrical symmetry of the isotropic
triangular lattice. This result is fully consistent with other works
for a $t$-$J$ model on a triangular lattice~\cite{ogata03,d+id}. The
continuous change of the relative phase $\theta$ is found by further
increasing the anisotropic parameter $t'/t$. For $t'/t > 1.3$, we
find that the value of $\theta$ saturates to $3\pi/4$.

\begin{figure}[t]
\includegraphics*[width=9cm,height=5.8cm,angle=0]{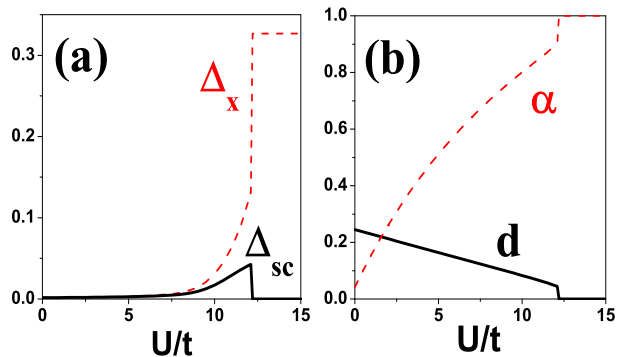}
\caption{Pairing amplitude $\Delta_x$ and its SC order parameter
$\Delta_x^{SC}$ (a), and electron double occupancy number $d$ and
the projection parameter $\alpha$ (b), as functions of $U/t$ for
$t'/t=1$.}
\end{figure}
Our detailed results for $t'/t=1$ are presented in Fig. 3. Here we
denote the SC order parameter as $\Delta_{SC}=\langle c_{i\downarrow
}c_{j\uparrow }-c_{i\uparrow}c_{j\downarrow }\rangle $. Within the
Gutzwiller approximation, $\Delta_{SC}$ can be expressed as
$\Delta_{SC}=g_{t}\Delta$. The effect of large $U$ may suppress the
double occupancy $d$ and stabilize the Mott insulating state. As
shown in Fig. 3(a), it is obvious that there exists a critical
$U_c$($\sim 12.2t$) to separate a SC state at a small $U$ from a
Mott insulator at a large $U$, and the transition is first order
with no co-existence of the two phases. At $U< U_c$, both $\Delta$
and $\Delta_{SC}$ increase monotonically as $U$ increases.
$\Delta_{SC}$ is smaller than $\Delta$ because of the prefactor $g_t
< 1$.  At the critical point, $\Delta$ increases to its maximum
while $\Delta_{SC}$ jumps to zero suddenly. At $U
> U_c$, the system becomes an insulator with $d=0$, or indicates
the full projection in the insulating phase . The projection
parameter $\alpha$ reaches its maximum in the SC phase followed by a
discontinuous jump to 1 at $U=U_c$. All these features suggest that
the first order transition is robust.

Another interesting feature reveals that in the SC phase just below
$U_c$, the double occupation number $d$ is quite small ($\sim
0.04$), while $\alpha$ is large ($\sim 0.91$). In accordance with
Laughlin's proposal of Gossamer superconductivity~\cite{laughlin},
very tiny superfluidity density appears in the SC state. It is
obvious that the SC state in this system can be well characterized
by the Gossamer superconductor. All the previous studies for less
frustrated system show the presence of appreciable (not so tiny)
superfluid density. In our present study, it is clear that the
effect of strong geometrically frustration may remarkably suppress
the superfluid density of the SC state~\cite{superfludity}.

\begin{figure}[b]
\includegraphics*[width=9cm,height=5.8cm]{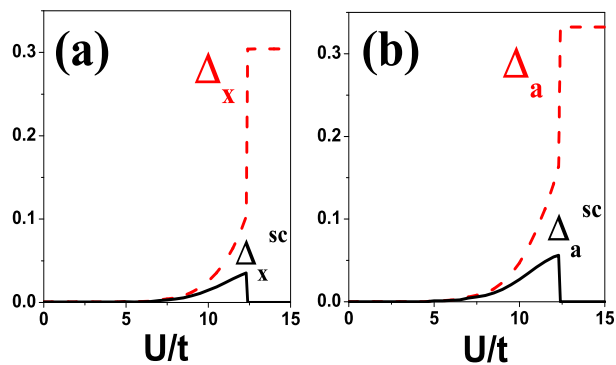}
\caption{Pairing amplitudes and its SC order parameter as functions
of $U/t$ for $t'/t=1.06$. (a) $\Delta_x$ and $\Delta_x^{SC}$. (b)
$\Delta_a$ and $\Delta_a^{SC}$.}
\end{figure}

Next we study the particular case for the compound
$\kappa$-(ET)$_2$Cu$_2$(CN)$_3$ which has a large degree of
geometrical frustration $t'/t \sim 1.06$. One expects the general
features are similar to the $t'/t=1$ case. As depicted in Fig. 4, we
notice two features which are in contrast to the $t'/t=1$ case. For
the anisotropic parameter $t'/t > 1$, the amplitude of $\Delta_{a}$
becomes larger than that of $\Delta_{x}$ while the SC order
parameter of $\Delta_a^{SC}$ is more pronounced than that of
$\Delta_x^{SC}$. In other words, the relative phase of SC order
parameter $\theta$ may deviate from the symmetrical value $2\pi/3$
and will be around 0.70$\pi$. Another distinctive feature shows that
the Mott transition point $U_c$ strongly depends on lattice
structure anisotropy $t^\prime/t$. Due to the competing nature
between Mott insulating phase and the SC phase, the increasing of
geometrical frustration $t^\prime/t$ may eventually destabilize the
Mott insulating phase which may lead to the Mott transition and
result in a SC phase. We find that the critical value increases
slightly from $U_c=12.2$ for $t'/t=1$ to $U_c=12.5$ for $t'/t=1.06$.
This result suggests that the increasing of $t'/t$ may effectively
enhance the critical value $U_c$.

\begin{figure}[t]
\centerline{\includegraphics[width=8.0cm]{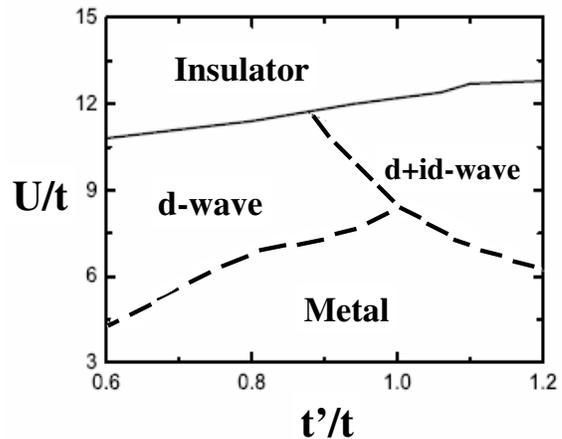}} \caption{Phase
diagram in parameter space $t'/t$ and $U/t$.}
\end{figure}
Fig. 5 displays the phase diagram in the parameter space  of $t'/t$
and $U/t$. It is technically difficult to determine precisely the
critical boundaries between various phases due to the presence of
degenerate states at half filling for highly frustrated system. In
practice, we define a paramagnetic metallic phase if both $\Delta_x$
and $\Delta_a$ are less than 0.01 and a $d+id$ state if $\Delta_{a}
> 0.01$. At very small $\Delta$, the energy discrepancy between
these phases becomes indistinguishable in our calculation. We use
dashed lines to represent the phase boundaries with very small
energy differences. The solid line denotes the first-order phase
boundary between SC phase and Mott insulating state. According to
our phase diagram, there exists four distinct phases. The system is
in the Mott insulating phase at large $U$ and small $t'$, the
paramagnetic metallic phase at small $U$ and large $t'$, and the two
possible SC phases at the intermediate parameter region. For small
$t'/t$, the phase diagram is consistent with our previous
study~\cite{gan05}. Note here that the time-reversal symmetry broken
$d+id$ state is preferable for large $t'/t$ while the pure
$d_{x^2-y^2}$-wave state is more stable for small $t'/t$. Since the
effect of pressure may increase $U/t$ and/or to increase $t'/t$, it
is quite plausible that a flow of parameters $U/t$ and $t'/t$ may
lead to a pairing symmetry transition from $d_{x^2+y^2}$ to $d+id$
state. In a very recent paper~\cite{Powell06}, the transition from
$d+id$ to $d$-wave is also indicated. Two recent
investigations~\cite{Liu05,Watanabe06} addressed the first-order
Mott transition by applying variational Monte Carlo method for the
layered organic superconductors by analyzing a two-dimensional
Hubbard model. By using a modified variational wavefunction with
doublon-holon binding factors, they found an unconventional SC
ground state for medium $U$, sandwiched between a normal metal at
weak coupling and a spin liquid at larger coupling. A first-order
Mott transition takes place at certain value of $U$. Our
renormalized mean field theory can take into account approximately
the effect of Gutzwiller projection and the results agree
qualitatively with theirs. It is well known that the
Brinkman-Rice~\cite{Brinkman} type metal-insulator transition for
Hubbard model belongs to the second-order. It seems to us that the
extra spin exchange term in the Hubbard-Heisenberg model makes much
difference and changes the Gossamer SC to Mott insulator transition
to be first-order.

In summary, we have utilized the Gutzwiller mean-field method to
study the $t$-$t'$-$J$-$J'$-$U$ model on a highly frustrated
anisotropic triangular lattice. We find that the pure $d_{x^2-y^2}$
state with nodal structure is stable if the geometrical frustration
parameter $t^\prime/t < (t^\prime/t)_c$ while the gapped $d+id$
state is more preferable for $t^\prime/t > (t^\prime/t)_c$ in the SC
phase. In the case of $t'/t=1$, the relative phase of the SC order
parameter is equal to $2\pi/3$ which reflects the geometrical
symmetry of isotropic triangular lattice. For a fixed $t^\prime/t$,
the ground state is found to be a paramagnetic metal at small
on-site Coulomb repulsion $U$; a Mott insulator at large $U$; and a
superconductor at intermediate $U$. This SC state can be well
described by the Gossamer superconductivity with tiny superfluid
density. The transition between SC and insulating phases belongs to
first order. %The effect of frustration makes the transitions more
%like a second-order type.
Our results may shed light on the
understanding of the exotic features of
$\kappa$-(ET)$_2$Cu$_2$(CN)$_3$, which undergoes a possible spin
liquid to superconductor transition at half filling under pressure.

This work is supported by the RGC grants of Hong Kong SAR
government, and seed funding grant from the University of Hong Kong.


\begin{thebibliography}{99}
\bibitem{footnote} In reference [26], one obtains the critical value $(t'/t)_c \sim
0.675$. The discrepancy is due to the fact that in their
calculation, the double occupation is fixed to zero and different
parameters as $J/t=0.3$ and $J^\prime/J=t^\prime/t$ are used.
\end{thebibliography}
\end{document}